\pdfoutput=1

\documentclass[11pt,a4paper]{article}
\usepackage{acl}

\usepackage[most]{tcolorbox}
\tcbuselibrary{skins,breakable}
\usepackage{wrapfig}

\usepackage{times}
\usepackage{latexsym}
\usepackage{graphicx}
\usepackage{booktabs}
\usepackage[T1]{fontenc}

\usepackage[utf8]{inputenc}

\usepackage{microtype}
\usepackage{inconsolata}

\usepackage{graphicx}
\usepackage{fancyhdr}
\usepackage{dblfloatfix} 
\usepackage{float}
\usepackage{pifont}      
\usepackage{array}       
\usepackage{makecell}    
\usepackage{ragged2e}    
\usepackage{rotating}
\usepackage{tabularx}
\usepackage{colortbl}
\usepackage{multirow}
\usepackage{enumerate}
\usepackage{longtable}
\usepackage{url}         
\usepackage{pdflscape}   
\usepackage{float}       
\usepackage{graphicx}
\usepackage{subcaption}

\newcolumntype{P}[1]{>{\centering\arraybackslash}p{#1}}
\newcolumntype{M}[1]{>{\centering\arraybackslash}m{#1}}
\newcolumntype{L}[1]{>{\raggedright\arraybackslash}m{#1}}
\newcolumntype{C}[1]{>{\centering\arraybackslash}m{#1}}

\usepackage{xcolor}
\usepackage{enumitem}
\setlist[description]{leftmargin=1.2em,labelsep=0.4em,itemsep=2pt,topsep=2pt,font=\bfseries}

\usepackage{xcolor}
\definecolor{darkgreen}{rgb}{0.0,0.4,0.0}
\newcommand{\codegreen}[1]{\textcolor{darkgreen}{\texttt{#1}}}

\title{APEX\textendash SWE}

\author{
    \textbf{Abhi Kottamasu}$^{1}$ \quad
    \textbf{Chirag Mahapatra}$^{1}$ \quad
    \textbf{Sam Lee}$^{2}$ \quad
    \textbf{Ben Pan}$^{2}$ \quad \\
    \textbf{Aakash Barthwal}$^{1}$ \quad
    \textbf{Akul Datta}$^{1}$ \quad
    \textbf{Anurag Gupta}$^{1}$ \quad
    \textbf{Pranav Mehta}$^{1}$ \quad
    \textbf{Ajay Arun}$^{1}$ \quad \\
    \textbf{Silas Alberti}$^{2}$ \quad
    \textbf{Adarsh Hiremath}$^{1}$ \quad
    \textbf{Brendan Foody}$^{1}$ \quad
    \textbf{Bertie Vidgen}$^{1}$\thanks{Email: apex@mercor.com} \\
    $^1$Mercor \quad $^2$Cognition \quad
}

\begin{document}
\pagestyle{fancy}
\fancyhf{}  

\fancyhead[L]{\includegraphics[width=\headwidth]{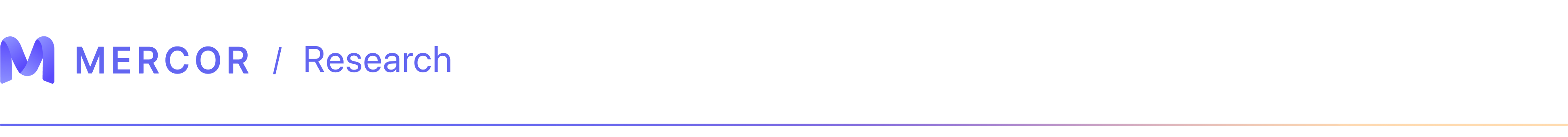}}

\fancyfoot[C]{\thepage}

\renewcommand{\headrulewidth}{0pt}

\fancypagestyle{plain}{
  \fancyhf{}
  \fancyhead[L]{\includegraphics[width=\headwidth]{images/banner.png}}
  \fancyfoot[C]{\thepage}
  \renewcommand{\headrulewidth}{0pt}
}

\maketitle
\thispagestyle{fancy}

\setlength{\parindent}{0pt}

\begin{abstract}
We introduce the AI Productivity Index for Software Engineering (APEX\textendash SWE), a benchmark for assessing whether frontier AI models can execute economically valuable software engineering work. Unlike existing evaluations that focus on narrow, well-defined tasks, APEX\textendash SWE assesses two novel task types that reflect real-world software engineering: (1) \mbox{\textbf{Integration}} tasks ($n=100$), which require constructing end-to-end systems across heterogeneous cloud primitives, business applications, and infrastructure-as-code services, and (2) \mbox{\textbf{Observability}} tasks ($n=100$), which require debugging production failures using telemetry signals such as logs and dashboards, as well as unstructured context. We evaluated eleven frontier models for the APEX\textendash SWE leaderboard. Claude Opus 4.6 leads the APEX\textendash SWE leaderboard with $40.5\%$ Pass@1, followed by Claude Opus 4.5 at $38.7\%$. Our analysis shows that strong performance is primarily driven by epistemic discipline, defined as the capacity to distinguish between assumptions and verified facts. It is often combined with systematic verification prior to acting. We open-source the APEX\textendash SWE evaluation harness and a dev set ($n=50$).
\end{abstract}

\begin{figure}[t]
\centering
\includegraphics[width=1\linewidth]{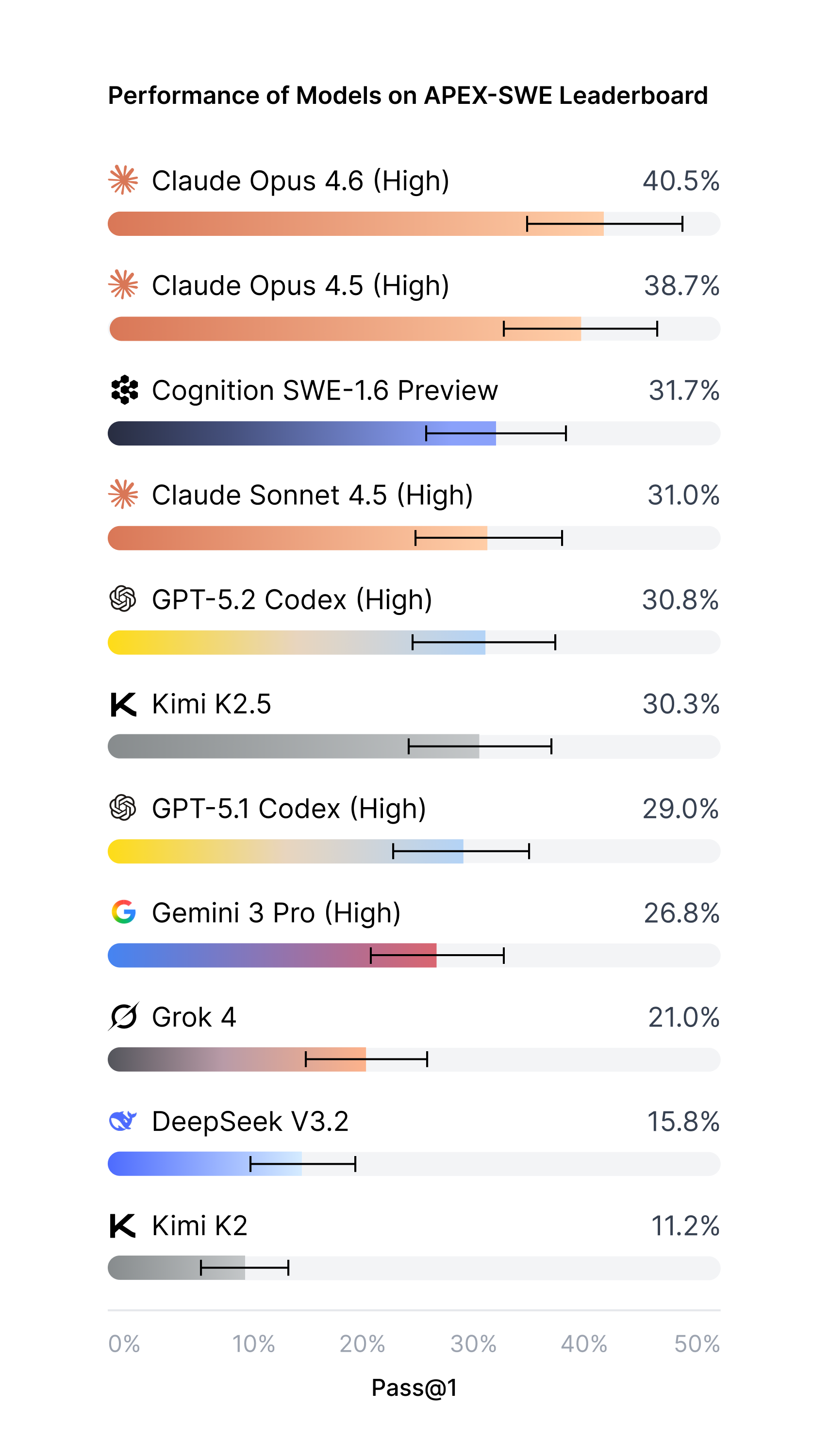}
\caption{Performance of models on APEX\textendash SWE using Pass@1. Thinking settings are in parentheses.}
\label{fig:benchmark_overall_performance}
\end{figure}

\section{Introduction}
Benchmarks for measuring real-world software engineering capability need to mirror actual development workflows. Yet industry data shows a fundamental mismatch between what existing benchmarks measure and what professional software engineers do: according to IDC, developers spend only $16\%$ of their time writing application code~\cite{idc2025developer}. The remaining $84\%$ involves CI/CD implementation, infrastructure monitoring, security, deployment, and debugging. Debugging alone costs the software industry \$312 billion annually ~\cite{cambridge2013debugging}.
\vspace{1em}

Integration work (connecting heterogeneous systems, configuring infrastructure, orchestrating cross-service workflows) is central to modern software development, as is Observability work (diagnosing production failures from logs, traces, and metrics). Yet no benchmark evaluates these capabilities systematically. SWE-bench Verified focuses exclusively on single-repository bug fixing, missing the cross-system integration and observability work that dominates real engineering practice. The benchmark has become saturated -- frontier models cluster around $80\%$ Pass@1, and OpenAI has declared the benchmark ``contaminated'' as models can reproduce original patches verbatim from task IDs alone~\cite{openai2025swebench}. 
\vspace{1em}

To assess models for real-world software engineering, we present \textbf{APEX\textendash SWE}, comprising Integration and Observability task types. We have released an open-source dev set on Hugging Face with a CC-BY license\footnote{\href{https://huggingface.co/datasets/mercor/APEX-SWE}{https://huggingface.co/datasets/mercor/APEX-SWE}} and our grading harness on GitHub.\footnote{\href{https://github.com/Mercor-Intelligence/apex-swe}{https://github.com/Mercor-Intelligence/apex-swe}} All of the models we evaluated fail to reliably solve the tasks in APEX\textendash SWE. Claude Opus 4.6 (Thinking=High) tops the APEX\textendash SWE leaderboard at $40.5\%$ on Pass@1, followed by Claude Opus 4.5 (Thinking=High) at $38.7\%$, as shown in Figure~\ref{fig:benchmark_overall_performance}. There is a substantial gap between current frontier systems and the reliability required for production-grade engineering.
\vspace{1em}

\begin{figure}[t]
\centering
\includegraphics[width=1\linewidth]{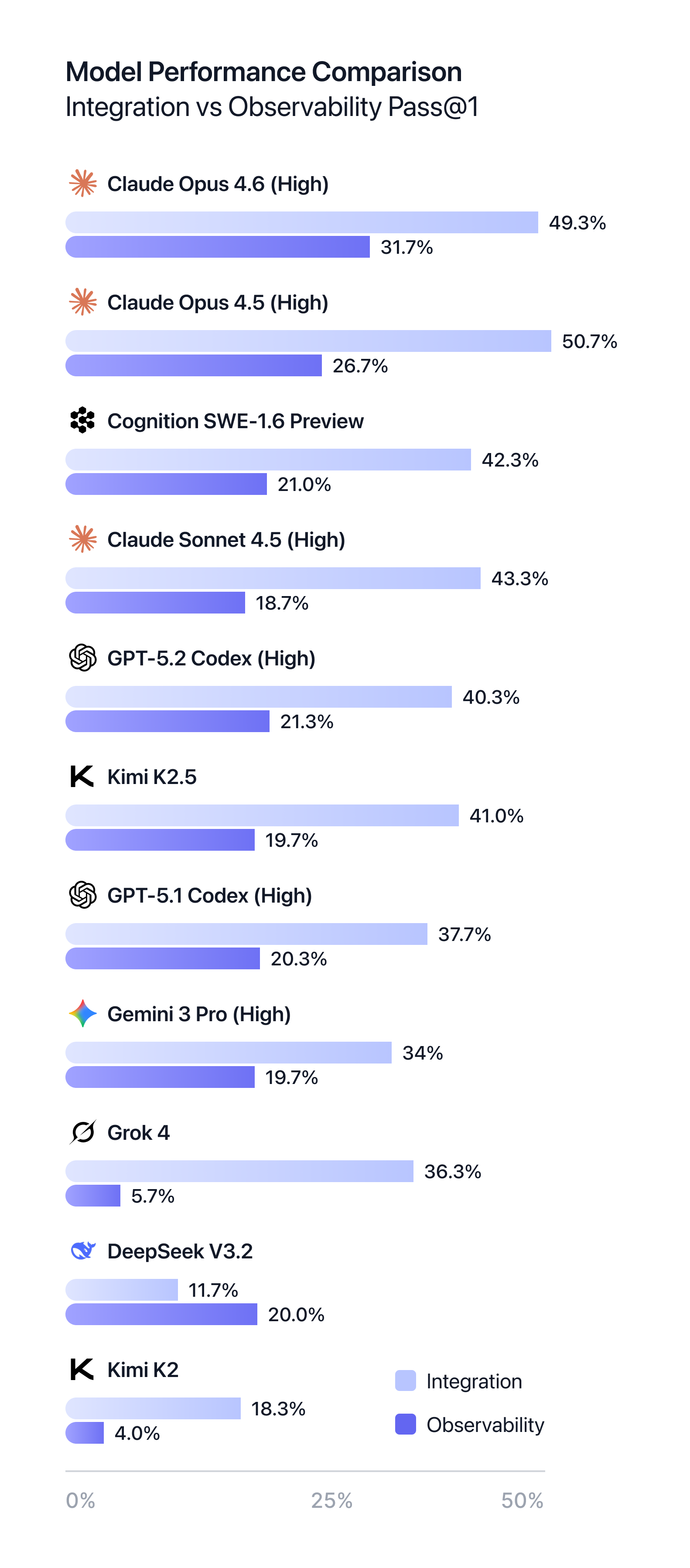}
\caption{Performance of models on APEX\textendash SWE Observability and APEX\textendash SWE Integration using Pass@1. Thinking settings are in parentheses.}
\label{fig:model-domain-score}
\end{figure}

As shown in Figure~\ref{fig:model-domain-score}, the best-performing model on Integration is Claude Opus 4.5 (Thinking=High) at $50.7\%$, followed by Claude Opus 4.6 (Thinking=High) at $49.3\%$, Claude Sonnet 4.5 (Thinking=High) at $43.3\%$, and Cognition SWE-1.6 Preview\footnote{Preview version of SWE-1.6, evaluated on 9 March 2026.} at $42.3\%$. Observability scores are lower overall, with Claude Opus 4.6 (Thinking=High) performing best at $31.7\%$ and most other models in the low-$20\%$ range or below. Successful models exhibit epistemic reasoning: they treat their code as a provisional hypothesis and iteratively validate it against the system's actual state.
\vspace{1em}

\begin{figure*}[t]
\centering
\includegraphics[width=1.05\linewidth]{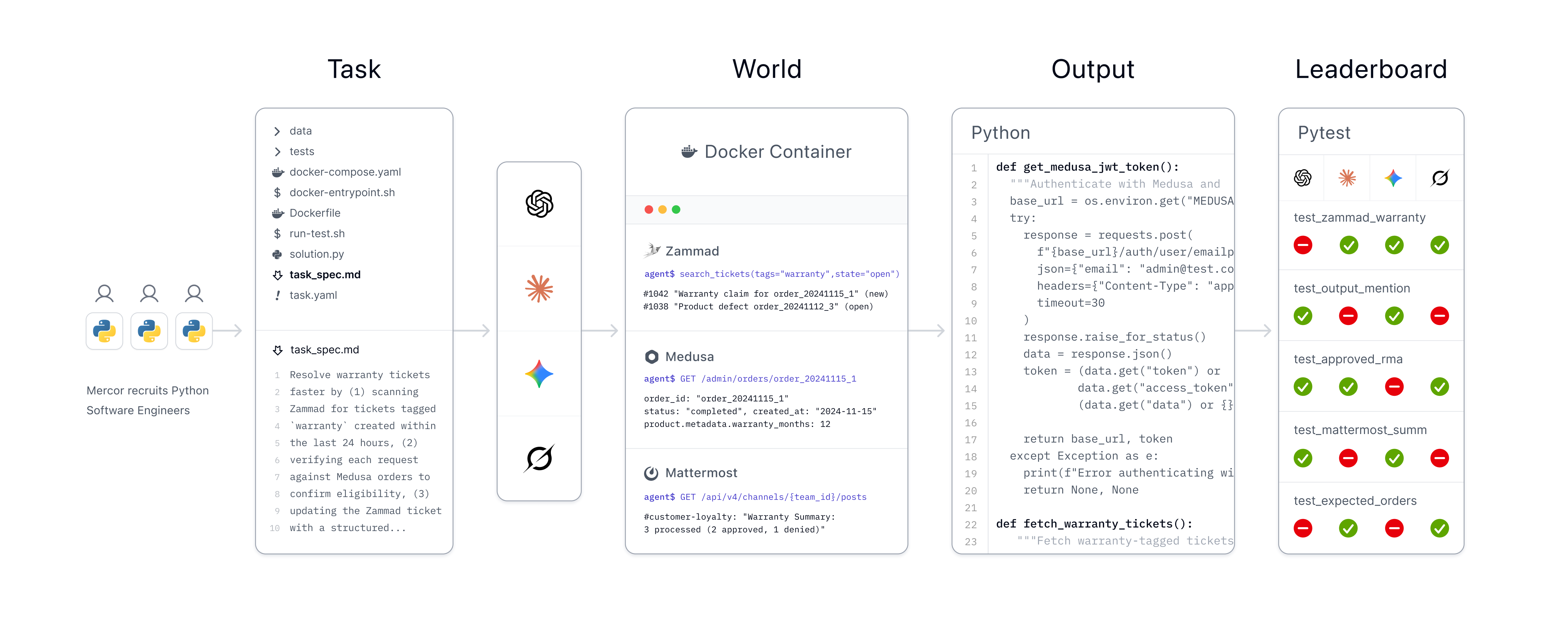}
\label{fig:swe-integration}
\caption{Production process for APEX\textendash SWE Integration.}
\label{fig:swe-integration-process}
\end{figure*}

\textbf{APEX\textendash SWE Integration} ($n=100$) evaluates a model's ability to orchestrate end-to-end workflows and synchronize data across heterogeneous services. Models are required to write application code, configure infrastructure, and deploy functioning services. The test stack includes cloud primitives (AWS LocalStack: S3, Lambda, DynamoDB, Kinesis) and production-grade business applications (EspoCRM, Medusa, Zammad, Plane).
\vspace{1em}

\textbf{APEX\textendash SWE Observability} ($n=100$) evaluates a model's ability to diagnose and remediate real-world production failures. Unlike conventional bug-fix benchmarks, models are not provided with failing unit tests. Instead, they must interrogate production logs (via Grafana/Loki), correlate evidence from developer chat discussions, and trace root causes through the codebase.
\vspace{1em}

\section{Experimental Setup}
\paragraph{Model Selection}
We evaluated eleven frontier models against APEX\textendash SWE: Claude Opus 4.5 (High), Claude Opus 4.6 (High), Claude Sonnet 4.5 (High), Cognition SWE-1.6 Preview (High), DeepSeek V3.2, Gemini 3 Pro (High), GPT\textendash 5.1 Codex (High), GPT\textendash 5.2 Codex (High), Grok 4, Kimi K2 Instruct, and Kimi K2.5. Thinking settings are set to High. In the Appendix, Table~\ref{tab:model_details} describes the model configurations.
\vspace{1em}

Models interact with the environment through a ReAct harness, which works in a persistent, multi-step loop. The model receives a system prompt containing task instructions and tool definitions. It iterates until it outputs a \codegreen{<task\_complete>} tag or reaches the wall-clock timeout of one hour. Models have access to three categories of tools: (1) Terminal (Bash execution via \codegreen{/bin/bash} in a persistent tmux session), (2) File Operations (with read/write access to the workspace), and (3) Model Context Protocol (MCP) Servers (i.e., API access to services such as Loki, Plane, Medusa).

\paragraph{Evals}
We assess models' outputs using Pass@1, defined as the average pass rate across three independent epochs (or ``runs''). For Integration, correctness is verified via a \codegreen{pytest} suite that interacts directly with service APIs. For Observability, correctness follows a \textsc{Fail\_to\_Pass} / \textsc{Pass\_to\_Pass} methodology inspired by SWE-bench~\citep{jimenez2024swebench}. We also report Pass@3, assessing if the model can achieve the correct outcome at least once over three epochs.

\begin{table}[b]
\centering
\small
\caption{Performance of models on APEX\textendash SWE Integration ($n=100$) with Pass@1 and Pass@3.}
\label{tab:integration_overallperformance}
\begin{tabular}{lcc}
\toprule
\textbf{Model}                                                  & \textbf{Pass@1}       & \textbf{Pass@3}        \\
\midrule
Claude Opus 4.5 (High)                                          & $\mathbf{50.7\%}$      & $\mathbf{58.0\%}$      \\
Claude Opus 4.6 (High)                                          & $49.3\%$               & $54.0\%$               \\
Claude Sonnet 4.5 (High)                                        & $43.3\%$               & $52.0\%$               \\
Cognition SWE-1.6 Preview (High)                                & $42.3\%$               & $52.0\%$               \\
DeepSeek V3.2                                                   & $11.7\%$               & $20.0\%$               \\
Gemini 3 Pro (High)                                             & $34.0\%$               & $39.0\%$               \\
GPT\textendash 5.1 Codex (High)                                 & $37.7\%$               & $47.0\%$               \\
GPT\textendash 5.2 Codex (High)                                 & $40.3\%$               & $51.0\%$               \\
Grok 4                                                          & $36.3\%$               & $49.0\%$               \\
Kimi K2 Instruct                                                & $18.3\%$               & $28.0\%$               \\
Kimi K2.5                                                       & $41.0\%$               & $51.0\%$               \\
\bottomrule
\end{tabular}
\end{table}

\paragraph{Qualitative Analyses}
We qualitatively analyze a sample of models' outputs, using LM-powered reviewing tools with extensive human oversight. 
\vspace{1em}

\section{APEX\textendash SWE Integration}
\subsection{Integration Tasks Data}
\paragraph{Task Sourcing} Tasks were created by software engineers with 3+ years of experience. Each case underwent a three-stage validation process. First, a check to verify that task prompts align with the source documents provided to the model. Second, test validation to ensure that the test suite faithfully evaluates the request without reward hacking. Third, creation of a gold standard output to ensure that the human-authored solution passes all tests with a 100\% score. The production process for SWE Integration tasks and our grading is shown in Figure~\ref{fig:swe-integration-process}.

\paragraph{Task Environment} All Integration tasks include an ephemeral PostgreSQL database and Plane, as well as six other services: LocalStack ($56\%$), which emulates AWS primitives (such as S3, Lambda, DynamoDB, and Kinesis), EspoCRM ($35\%$), MailHog ($33\%$) for SMTP testing, Mattermost ($32\%$), Medusa ($31\%$) for e-commerce, and Zammad ($26\%$).

\paragraph{Authentication \& Security} Tasks have authentication schemes to test models' ability to handle credential management, including Basic Auth for EspoCRM and Zammad, JWT tokens for Medusa, IAM policies and STS credentials for LocalStack, API keys for various webhooks, and PostgreSQL connection strings. Models must read credentials from environment variables rather than hardcoding values, enforcing production-grade security.
\vspace{1em}

\begin{figure*}[t]
\centering
\includegraphics[height=0.45\textheight, keepaspectratio]{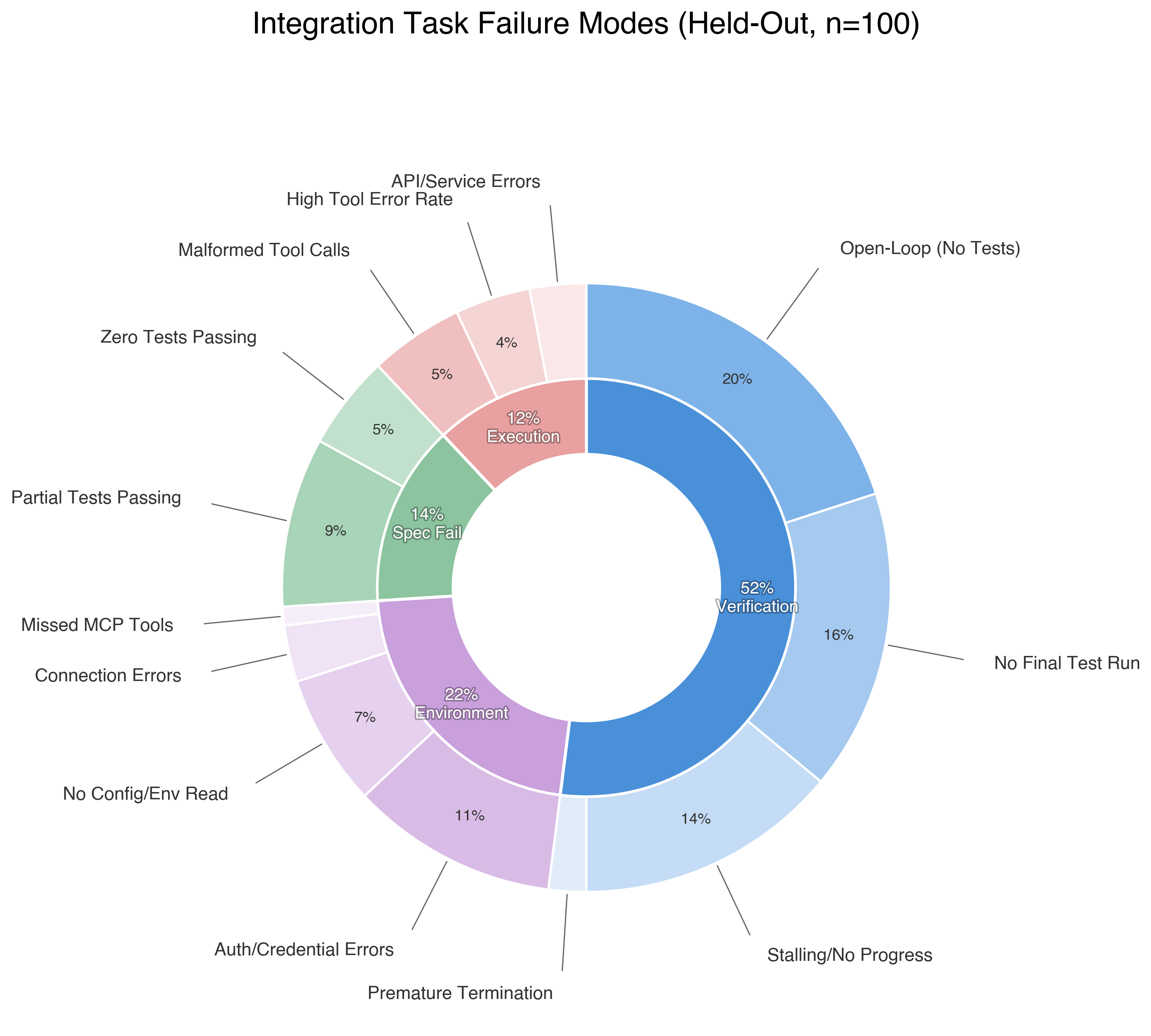}
\caption{Integration task failure modes, aggregated across all models (held-out, $n=100$). Inner ring shows Tier 1 root causes; outer ring shows Tier 2 sub-categories.}
\label{fig:integration-failure-sunburst}
\end{figure*}

\subsection{Performance on APEX\textendash SWE Integration}
Model performance on the Integration tasks is summarized in Table~\ref{tab:integration_overallperformance}. Claude Opus 4.5 leads in single-shot performance with $50.7\%$ Pass@1, followed by Claude Opus 4.6 at $49.3\%$, Claude Sonnet 4.5 at $43.3\%$, and Cognition SWE-1.6 Preview at $42.3\%$. A competitive middle tier includes Kimi~K2.5 ($41.0\%$ Pass@1), GPT\textendash5.2 Codex ($40.3\%$), and GPT\textendash5.1 Codex ($37.7\%$). Grok~4 achieves $36.3\%$, followed by Gemini~3~Pro at $34.0\%$. Kimi~K2 Instruct achieves $18.3\%$ Pass@1, while DeepSeek~V3.2 trails at $11.7\%$, marking a substantial gap between frontier and lower-tier models. The Claude family also demonstrates higher peak potential, with Opus~4.5, Opus~4.6, and Sonnet~4.5 achieving the strongest Pass@3 results (58\%, 54\%, and 52\%, respectively).

\begin{table*}[b]
\centering
\small
\caption{Average number of episodes on APEX\textendash SWE Integration ($n=100$). Results are shown for all tasks, and split by Successful tasks and Failed tasks.}
\label{tab:average_time_taken_models_and}
\begin{tabular}{l|ccc}
\toprule
\textbf{Model}
& \makecell{\textbf{Number of}\\ \textbf{Episodes (All)}}
& \makecell{\textbf{Number of}\\ \textbf{Episodes (Success)}}
& \makecell{\textbf{Number of}\\ \textbf{Episodes (Fail)}} \\
\midrule
Claude Opus 4.5 (High)                                          & $60.6$ & $51.0$ & $70.7$ \\
Claude Opus 4.6 (High)                                          & $48.6$ & $38.5$ & $58.9$ \\
Claude Sonnet 4.5 (High)                                        & $66.7$ & $59.3$ & $72.5$ \\
Cognition SWE-1.6 Preview (High)                                & $85.2$ & $71.4$ & $93.7$ \\
DeepSeek V3.2                                                   & $60.0$ & $58.6$ & $60.2$ \\
Gemini 3 Pro (High)                                             & $37.0$ & $32.6$ & $39.4$ \\
GPT\textendash 5.1 Codex (High)                                 & $61.3$ & $44.3$ & $71.8$ \\
GPT\textendash 5.2 Codex (High)                                 & $42.9$ & $35.7$ & $47.8$ \\
Grok 4                                                          & $51.7$ & $39.2$ & $59.0$ \\
Kimi K2 Instruct                                                & $19.2$ & $17.4$ & $19.9$ \\
Kimi K2.5                                                       & $55.0$ & $46.1$ & $61.3$ \\
\midrule
\textbf{Average}                                                & $53.5$ & $44.9$ & $59.6$ \\
\bottomrule
\end{tabular}
\end{table*}

\subsection{Success Analysis}
We qualitatively reviewed the successful Integration runs. We find that success is not determined by raw coding capability; instead, it is driven by epistemic discipline -- an agent's ability to distinguish between assumptions and facts, and its willingness to verify the former before acting. High-performing models often exhibit a three-phase workflow: systematic exploration, explicit specification extraction, and closed-loop verification.

\paragraph{Problem Space Exploration}
Successful agents prioritize building context over taking immediate action, investing early episodes in defining their strategy. Rather than immediately coding, they inspect data structures and query tools. In $70.5\%$ of passing trajectories, models explicitly read environment configuration files before attempting implementation, and $95.0\%$ queried external APIs to discover available operations. In one task, Claude Opus 4.5 explicitly identified that it needed to retrieve full issue details via MCP before implementing the fix. It executed shell commands to verify input schemas rather than inferring them, minimizing the risk of hallucination.

\paragraph{Task Specification}
Once the environment is modeled, successful agents parse unstructured descriptions into lists of hard constraints. They treat resource names and paths as explicit specifications. In one instance, an agent produced a numbered list of requirements (e.g., ``Create bucket daily-snapshots-bucket... Key format daily/YYYY-MM-DD/users.csv'') before generating code.

\paragraph{Verification \& Self-Diagnosis}
A key feature of successful agents is that they do not accept code generation on its own as evidence of task completion. These agents explicitly verify specifications by mapping outputs back to requirements. In one task, Claude Opus 4.5 ran a script to empirically confirm file existence, then walked through a self-generated checklist to verify that uploads succeeded and cron entries were valid. This self-verification step prevents false positive completion.

\subsection{Failure Analysis}
We qualitatively reviewed the Integration tasks where the agent did not execute the task. Figure~\ref{fig:integration-failure-sunburst} shows the distribution of failure modes aggregated across all models. Insufficient Verification accounts for $52\%$ of failures, followed by Bad Environment Understanding ($22\%$), Specification Non-Compliance ($14\%$), and Execution Failure ($12\%$).

\paragraph{Insufficient Verification (52\%)}
The dominant failure mode is open-loop execution, where models assume code generation equals task completion. Within this category, $20\%$ of signals stem from models that never run test scripts, $16\%$ from models that skip final verification, and $14\%$ from stalling without progress. In one task, GPT\textendash 5.2 Codex implemented a script that correctly identified low-stock products and sent an alert email, passing five of seven tests. The failures were due to the script reporting three products when the actual count was six. The model never verified that its output matched the required state.

\paragraph{Bad Environment Understanding (22\%)}
A recurring pattern is agents bypassing provided MCP tools in favor of raw HTTP calls. In one task, GPT\textendash 5.1 Codex was asked to query Plane for completed issues, build a JSON report, and upload it to LocalStack S3. The model spent over $60\%$ of its 83-episode trajectory fighting 403 Forbidden errors on the Plane API, trying wrong workspace slugs, and debugging empty responses. It eventually created a script that reported ``0 completed issues'' because it could not authenticate correctly. All 6 tests failed. The MCP tools would have handled authentication automatically, but the model chose to construct raw HTTP calls instead -- trading familiar interfaces for unnecessary complexity.

\paragraph{Specification Non-Compliance (14\%)}
Specification errors occur when models implement core functionality but miss exact format requirements or defensive coding expectations. In one task with an automated welcome email campaign, GPT\textendash 5.2 Codex was asked to fetch new EspoCRM contacts, send personalized emails via MailHog SMTP, and update contact descriptions with timestamps. The model created a working script, ran it, and verified that 7 welcome emails were sent -- passing 6 of 8 tests. However, it failed the exact email format test (subject and body did not match the required template) and the email count test.
\vspace{1em}

\paragraph{Execution Failure (12\%)}
Malformed tool calls account for the majority of these failures, typically arising from broken JSON, incorrect argument formats, or attempts to invoke nonexistent functions. Grok~4 and GPT\textendash5.1 Codex are disproportionately affected. In particular, Grok~4 exhibits malformed tool calls in $25\%$ of its failing trajectories, making this the dominant failure mode for the model.
\vspace{1em}

Failing trajectories share a common characteristic: insufficient verification. Successful models treat task requirements as constraints to be checked, not guidelines to be approximated. The difference is epistemic discipline -- the willingness to verify assumptions against observable state rather than proceeding from plausible inferences.

\subsection{Number of Episodes to Complete Integration Tasks}
Models typically use fewer episodes on successful tasks compared with unsuccessful tasks ($44.9$ vs $59.6$), as shown in Table~\ref{tab:average_time_taken_models_and}. Claude Opus 4.6 is efficient, achieving strong performance ($49.3\%$ Pass@1) with the lowest overall turn count among top-performing models ($48.6$ avg). Its episode efficiency stems from an epistemic workflow: focused context gathering, immediate implementation, and verification. In contrast, Kimi K2 Instruct has the lowest turn count ($19.2$) but also lower performance ($18.3\%$). This speed stems from the model acting prematurely without sufficient context, where it starts executing without verifying context and working iteratively.
\vspace{1em}

\begin{figure*}[t]
\centering
\includegraphics[width=1.05\linewidth]{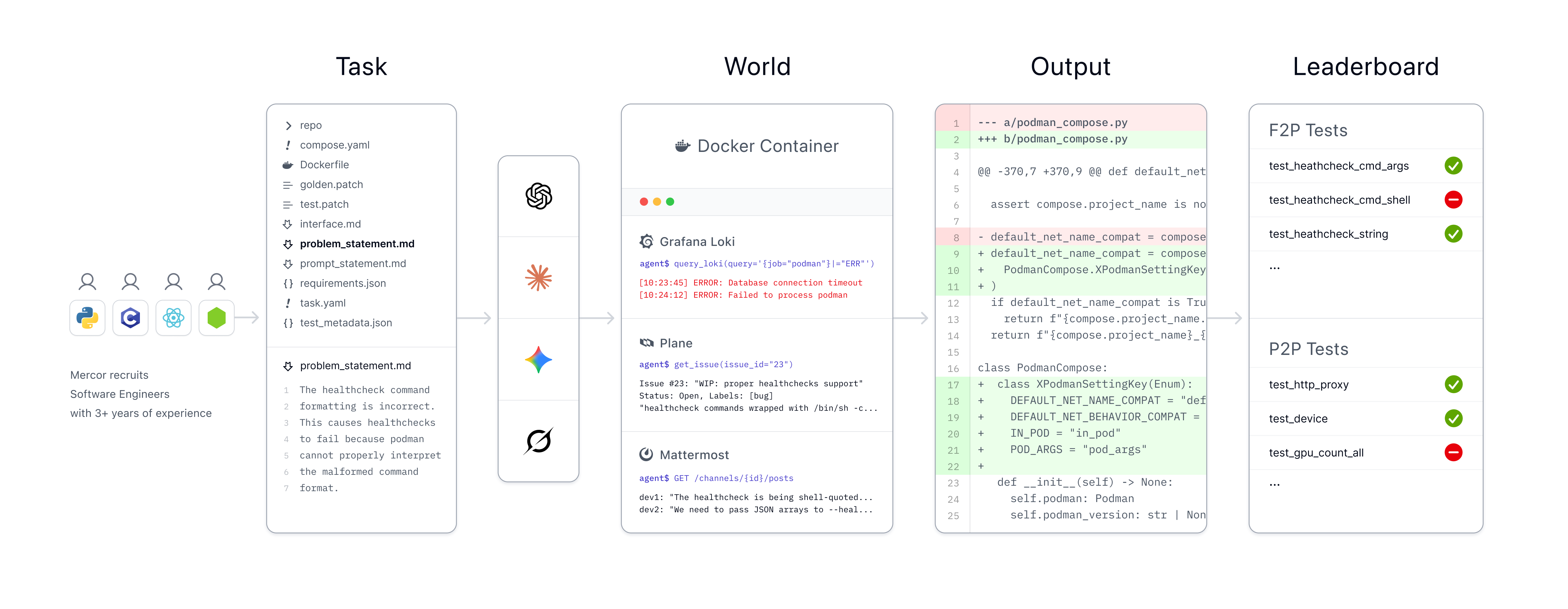}
\caption{Production process for APEX\textendash SWE Observability.}
\label{fig:swe-observability-process}
\end{figure*}

\section{APEX\textendash SWE Observability}

\subsection{Observability Tasks Data}
\paragraph{Task Sourcing} Tasks require agents to diagnose production failures from logs (Grafana/Loki) and chat context (Mattermost, Discord). No failing unit tests are provided; models must interrogate the environment to diagnose the failures. The tasks are derived from real-world GitHub Issue -- PR pairs, sourced from repositories with at least 350 stars. We filtered for complexity, selecting only patches with at least 100 lines of code impacting at least three files. Tasks were also evaluated for test coverage and brittleness, defined as whether tests are implementation-agnostic. Software engineers  with 3+ years of experience converted open-source tasks into benchmark tasks. The production process for SWE Observability tasks, as well as our grading process, is shown in Figure~\ref{fig:swe-observability-process}.

\paragraph{Task Environment} Each task deploys a containerized environment orchestrating five services: a client workspace, Loki and Promtail for log aggregation, Grafana for visualization, and Plane/Mattermost for ticket and chat context. Engineers scripted synthetic logs (500--1{,}000 lines of normal operation mixed with 10--20 lines of bug symptoms) and chat history to replicate a production failure. The model is provided with a summary of the user issue, an \codegreen{interface.md} file describing relevant repository functions, and specifications for available Model Context Protocol (MCP)~\cite{anthropic2024mcp} tools. Chat data are pulled from public developer discussions about the repo, including GitHub Discussions and Gitter, and GitHub issue data are taken directly from the repo. Observability tasks additionally include several key files: \codegreen{docker-compose.yaml} and \codegreen{Dockerfile} for service orchestration and defining containers, a \codegreen{task.yaml} containing task metadata, a data/ directory with seed data for all services, and two patches: \codegreen{test.patch} and \codegreen{golden.patch}, which contain test fixtures applied at evaluation time and the reference solution for validation respectively.

\paragraph{Programming Language} Observability tasks are across five languages: Go (30\%), Python (25\%), TypeScript (25\%), Java (10\%), and C++ (10\%).
\vspace{1em}

\begin{table}[t]
\centering
\small
\caption{Performance of models on APEX\textendash SWE Observability ($n=100$) with Pass@1 and Pass@3.}
\label{tab:observability_overallperformance}
\begin{tabular}{lcc}
\toprule
\textbf{Model} & \textbf{Pass@1} & \textbf{Pass@3} \\
\midrule
Claude Opus 4.5 (High)                                          & $26.7\%$               & $34.0\%$               \\
Claude Opus 4.6 (High)                                          & $\mathbf{31.7\%}$      & $\mathbf{39.0\%}$      \\
Claude Sonnet 4.5 (High)                                        & $18.7\%$               & $22.0\%$               \\
Cognition SWE-1.6 Preview (High)                                & $21.0\%$               & $31.0\%$               \\
DeepSeek V3.2                                                   & $20.0\%$               & $22.0\%$               \\
Gemini 3 Pro (High)                                             & $19.7\%$               & $31.0\%$               \\
GPT\textendash 5.1 Codex (High)                                 & $20.3\%$               & $31.0\%$               \\
GPT\textendash 5.2 Codex (High)                                 & $21.3\%$               & $29.0\%$               \\
Grok 4                                                          & $5.7\%$                & $9.0\%$                \\
Kimi K2 Instruct                                                & $4.0\%$                & $9.0\%$                \\
Kimi K2.5                                                       & $19.7\%$               & $26.0\%$               \\
\bottomrule
\end{tabular}
\end{table}

\subsection{Performance on APEX\textendash SWE Observability}
We evaluate eleven frontier models on 100 held-out Observability tasks. Table~\ref{tab:observability_overallperformance} presents the main results for Observability. Claude Opus 4.6 leads at $31.7\%$ Pass@1, followed by Claude Opus 4.5 at $26.7\%$ and GPT\textendash 5.2 Codex at $21.3\%$. A middle tier clusters tightly: Cognition SWE-1.6 Preview at $21.0\%$, GPT\textendash 5.1 Codex at $20.3\%$, DeepSeek V3.2 at $20.0\%$, Kimi K2.5 and Gemini 3 Pro at $19.7\%$, and Claude Sonnet 4.5 at $18.7\%$. The bottom tier drops sharply -- Grok 4 at $5.7\%$, Kimi K2 Instruct at $4.0\%$. Pass@3 scores show models have substantial headroom. Claude Opus 4.6 gains the most, jumping from $31.7\%$ to $39.0\%$ -- a 7.3 percentage point improvement with additional attempts. In contrast, Claude Sonnet 4.5 gains only 3.3 points ($18.7\%$ to $22.0\%$).

\subsection{Success Analysis for Observability tasks}
Success analysis on APEX\textendash SWE Observability tasks shows similar findings as the Integration tasks, with strong performance also driven by epistemic discipline. However, here, such discipline changes from verifying specifications to understanding the state of the system before attempting to repair it.

\paragraph{Iterative Debugging}
High-performing models treat log analysis as a search problem. They issue broad queries first, observe the result volume, and progressively narrow with filters. In one task, Claude Opus 4.5 issued an initial Loki query that returned 15{,}729 characters. Rather than parsing this output directly, the model refined its query with a LogQL filter, reducing output to 2{,}354 characters. Subsequent refinements brought the result down further -- to 264 characters -- until the relevant error was isolated. This behavior is critical for Observability because log volumes in production systems routinely exceed context window limits; models that query logs in one shot often trigger truncation warnings and proceed with incomplete data.

\paragraph{Multi-Source Triangulation}
Successful agents query multiple observability sources -- Loki logs, Mattermost discussions, and Plane issues -- before forming a diagnosis. In one task, Claude Opus 4.5 queried Loki first to identify error patterns, then searched Mattermost for developer discussion about the symptom, and finally retrieved the linked Plane issue for specification context. This triangulation appeared across $100\%$ of analyzed passing trajectories (all used at least one MCP tool beyond Loki). The pattern matters because Observability tasks embed diagnostic signals across sources: logs show what failed, chat shows what developers suspected, and issues show what behavior was originally intended. Models that skip sources miss the constraints that distinguish correct fixes from superficially similar ones.

\paragraph{System Exploration}
Successful agents explore the environment, reading an average of 11.3 files before attempting a fix. In one task involving a C++ IR remote library, Claude Opus 4.6 traced execution across 23 files before identifying the root cause in a toggle mechanism. The model read header files, test fixtures, and protocol implementations to build a complete picture of the state machine.
\vspace{1em}

\begin{table*}[b]
\centering
\scriptsize
\setlength{\tabcolsep}{3pt}
\caption{Performance of models on APEX\textendash SWE Observability tasks ($n=100$), split by programming language.}
\label{tab:observability_languageperformance}
\makebox[\textwidth][c]{
\resizebox{1.1\textwidth}{!}{
\begin{tabular}{l|c|ccccccccccc}
\toprule
\textbf{Language}
& \makecell{\textbf{Mean} \\ \textbf{Pass@1}}
& \makecell{\textbf{Claude Opus} \\ \textbf{4.5 (High)}}
& \makecell{\textbf{Claude Opus} \\ \textbf{4.6 (High)}}
& \makecell{\textbf{Claude Sonnet} \\ \textbf{4.5 (High)}}
& \makecell{\textbf{Cognition} \\ \textbf{SWE-1.6 Preview (High)}}
& \makecell{\textbf{DeepSeek} \\ \textbf{V3.2}}
& \makecell{\textbf{Gemini 3 Pro} \\ \textbf{(High)}}
& \makecell{\textbf{GPT\textendash 5.1} \\ \textbf{Codex (High)}}
& \makecell{\textbf{GPT\textendash 5.2} \\ \textbf{Codex (High)}}
& \textbf{Grok 4}
& \makecell{\textbf{Kimi K2} \\ \textbf{Instruct}}
& \textbf{Kimi K2.5} \\
\midrule
Python ($n=25$)     & $27.3\%$ & $32.0\%$ & $\mathbf{45.3\%}$ & $25.3\%$ & $40.0\%$ & $34.7\%$  & $24.0\%$ & $30.7\%$ & $30.7\%$ & $4.0\%$  & $2.7\%$  & $30.7\%$ \\
Go ($n=30$)         & $20.4\%$ & $\mathbf{36.7\%}$ & $30.0\%$ & $23.3\%$ & $18.9\%$ & $21.1\%$  & $22.2\%$ & $17.8\%$ & $22.2\%$ & $5.6\%$  & $4.4\%$  & $22.2\%$ \\
C++ ($n=10$)        & $20.0\%$ & $30.0\%$ & $\mathbf{43.3\%}$ & $10.0\%$ & $23.3\%$ & $13.3\%$  & $20.0\%$ & $20.0\%$ & $23.3\%$ & $13.3\%$ & $0.0\%$  & $23.3\%$ \\
TypeScript ($n=25$) & $12.1\%$ & $12.0\%$ & $16.0\%$ & $13.3\%$ & $8.0\%$ & $12.0\%$  & $\mathbf{18.7\%}$ & $14.7\%$ & $14.7\%$ & $6.7\%$  & $8.0\%$  & $9.3\%$ \\
Java ($n=10$)       & $10.0\%$ & $16.7\%$ & $\mathbf{30.0\%}$ & $10.0\%$ & $10.0\%$ & $6.7\%$ & $3.3\%$  & $16.7\%$ & $10.0\%$ & $0.0\%$  & $0.0\%$  & $6.7\%$ \\
\bottomrule
\end{tabular}
}}
\end{table*}

\subsection{Failure Analysis for Observability tasks}
\begin{figure*}[t]
\centering
\includegraphics[height=0.45\textheight, keepaspectratio]{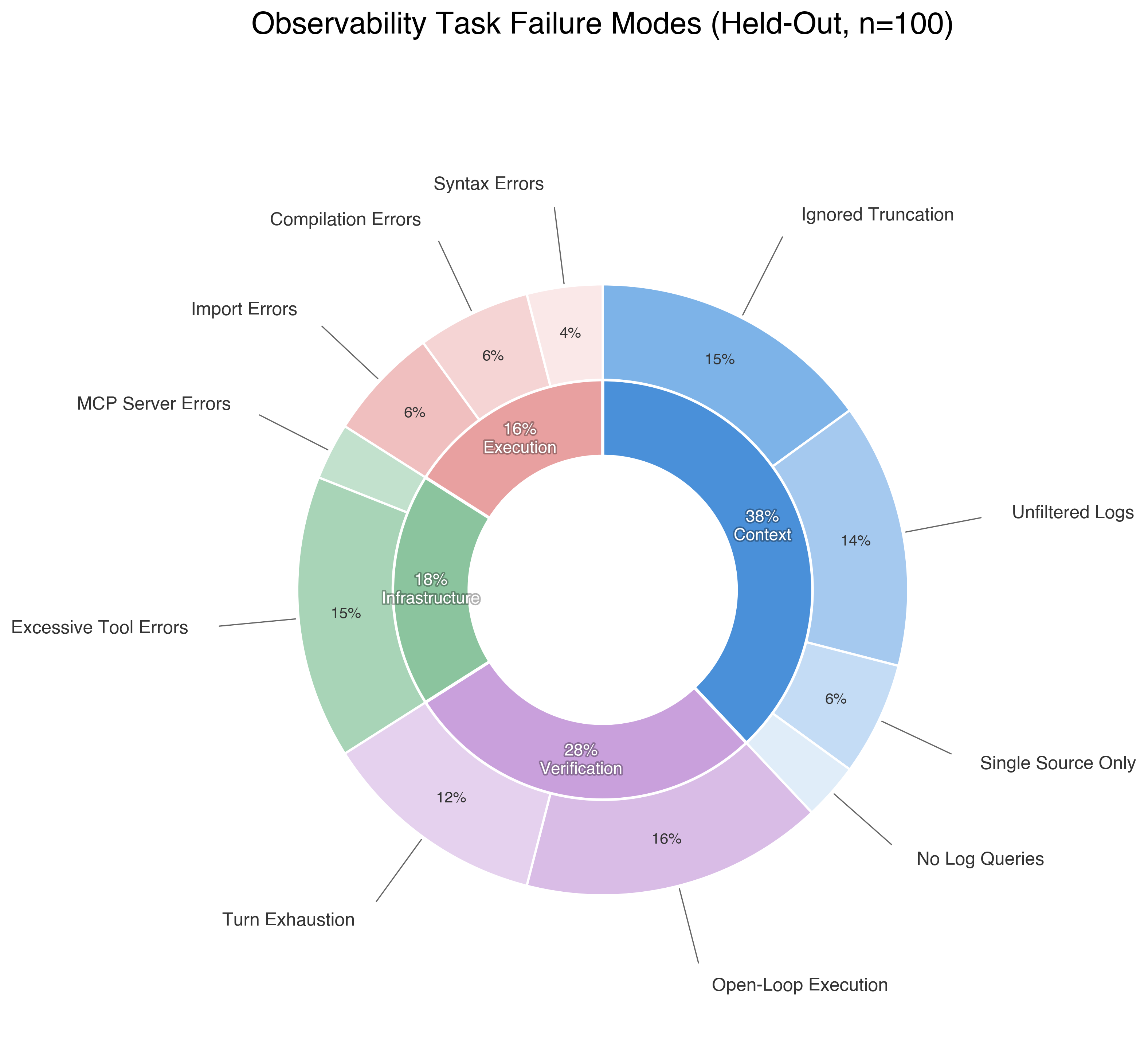}
\caption{Observability task failure modes aggregated across all models (held-out, $n=100$). Inner ring shows Tier 1 root causes; outer ring shows Tier 2 sub-categories.}
\label{fig:observability-failure-sunburst}
\end{figure*}

Figure~\ref{fig:observability-failure-sunburst} shows the distribution of failure modes for Observability tasks. Bad Context Handling dominates at $38\%$ of failure signals, followed by Insufficient Verification ($28\%$), Infrastructure Failure ($18\%$), and Execution Failure ($16\%$).

\paragraph{Bad Context Handling (38\%)}
The dominant failure mode involves poor log analysis strategies. Ignored truncation warnings account for $15\%$ of signals, and unfiltered log queries account for $14\%$. In the same task where Claude Opus 4.5 succeeded through iterative refinement, DeepSeek V3.2 issued a single, unfiltered query that returned 16{,}515 characters, immediately triggering a truncation warning. It then ignored the warning, did not refine the query, and patched files based on incomplete data. Single-source reliance ($6\%$) compounds this pattern -- models that query only Loki logs miss diagnostic signals embedded in Mattermost discussions or Plane issues.

\paragraph{Insufficient Verification (28\%)}
Open-loop execution accounts for $16\%$ of signals, while turn exhaustion accounts for $12\%$. Grok 4 is the worst at self-verification, with $86\%$ of trajectories ending with code edits without any verification.

\paragraph{Infrastructure Failure (18\%)}
Excessive Tool Failures account for $16\%$ of signals and captures trajectories where repeated tool errors cascade into task failure. Two models illustrate how specific tool weaknesses become fatal bottlenecks. Grok~4 fails more than a third of its \codegreen{str\_replace} calls on the first attempt, requiring multiple retries for the majority of its edits. DeepSeek~V3.2 shows a different failure signature: nearly half its \codegreen{view\_file} calls fail due to persistent argument format errors. In both cases, the failure mode is not any single error but the compounding effect: each failed call consumes episodes and triggers retry loops until the agent fails to complete the diagnostic task.                                       

\paragraph{Execution Failure (16\%)}
While some models fail because they cannot understand the problem, others fail because they cannot implement the solution. Import errors ($6\%$), compilation errors ($6\%$), and syntax errors ($4\%$) constitute this category. Kimi K2 and Gemini 3 Pro have the highest rates of execution failures. A recurring sub-pattern is agents importing a library for log analysis or data transformation that is not installed, hitting \codegreen{ModuleNotFoundError}, and not recovering.

\begin{figure*}[t]
\centering
\includegraphics[width=1.05\linewidth]{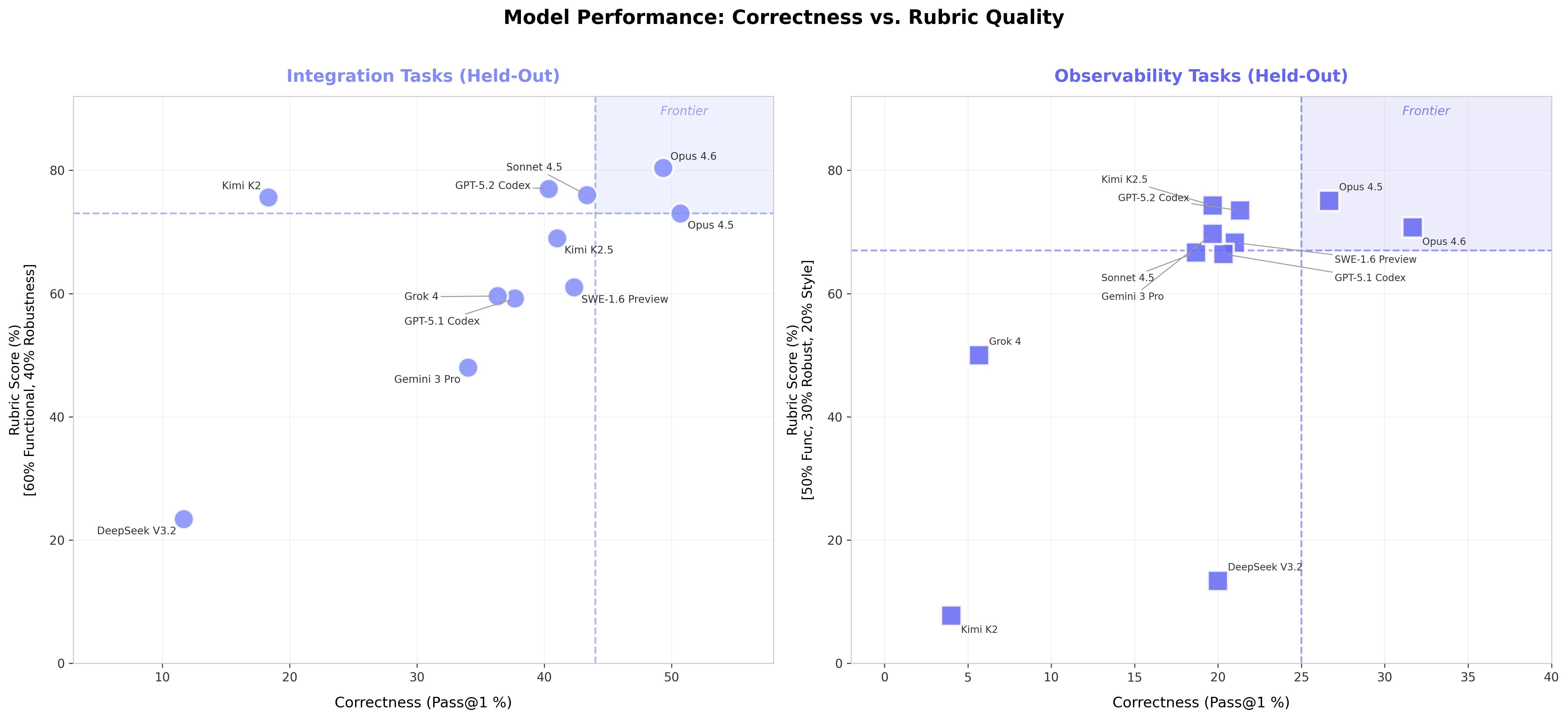}
\caption{Correctness (Pass@1) vs. Rubric Quality for Integration and Observability tasks on the held-out set. }
\label{fig:pareto_correctness_rubric}
\end{figure*}

\subsection{Performance by Language}
We segment model performance across five programming languages to isolate how build environments and type systems affect diagnostic ability, as shown in Table~\ref{tab:observability_languageperformance}. Models perform best on Python tasks ($27.3\%$ mean Pass@1), followed by Go ($20.4\%$) and C++ ($20.0\%$). TypeScript and Java cluster at $12.1$--$10.0\%$, roughly half the Python rate. The gap between Python and Java -- 17.3 percentage points -- reflects how runtime feedback availability shapes diagnostic success.

\paragraph{Python}
Python's mean Pass@1 is $27.3\%$, the highest on average. Its permissive runtime supports self-correction. Agents execute imperfect code, observe error feedback, and refine their approach.

\paragraph{Go}
Go's $20.4\%$ mean Pass@1 reflects its explicit error handling patterns. Go's compile-time feedback and structured error returns align well with the verify-then-iterate workflow that characterizes successful Observability trajectories.

\paragraph{C++}
C++ achieves $20.0\%$ mean Pass@1 despite compilation requirements. The C++ tasks ($n=10$) involve embedded systems code with relatively self-contained structure. Success on C++ requires reasoning about header dependencies and build config.

\paragraph{TypeScript and Java}
TypeScript ($12.1\%$) and Java ($10.0\%$) are similarly difficult. TypeScript's strict type checking (\codegreen{strictNullChecks}, complex module resolution) limits the feedback loop during debugging. In contrast, Java tasks use the Halo platform with Spring WebFlux; models struggle with reactive programming patterns that deviate from synchronous request-response flows.

\section{Rubric Analysis}
Beyond Pass@1, APEX\textendash SWE evaluates engineering quality through rubrics. Rubrics can be used to score qualitative and open-ended aspects of outputs through self-contained, objective statements that describe important attributes of a high-quality response \citet{arora2025healthbenchevaluatinglargelanguage}. Pass@1 measures binary task completion; rubrics measure the quality of the process and reward nice-to-have attributes of model outputs -- suich as whether the code handles edge cases, follows specifications, and adheres to conventions. \textbf{Rubric scores are not used for the APEX\textendash SWE leaderboard}, which relies exclusively on Pass@1. Rubrics are task-specific and created by experience engineers who worked on the task. They are each graded independently by an LM judge (Gemini 3 Pro, Temperature~0.1, Thinking=High). The judge evaluates the final patch and execution logs against these task-specific criteria. We define three rubric categories.

\paragraph{Functional Criteria}
Functional criteria assess whether implementations satisfy core behavioral requirements (e.g., ``Mutex must be locked before reading current values''), ensuring that the system behaves correctly under expected operating conditions.

\paragraph{Robustness Criteria}
Robustness criteria evaluate defensive coding practices: exception handling, input validation, edge case coverage, and graceful failure modes under unexpected conditions.

\paragraph{Style Criteria}
Style criteria, evaluated only for Observability tasks, assess documentation quality, code organization, and adherence to language-specific conventions. Observability tasks involve patching existing codebases where conformance to repository conventions matters for maintainability.

\subsection{Correctness vs. Rubric Quality}
Figure~\ref{fig:pareto_correctness_rubric} shows that correctness and quality do not always align. The frontier -- models achieving both high Pass@1 and high rubric scores -- is occupied by Claude Opus 4.5 and 4.6 on both task types. These frontier models combine task completion with engineering discipline: they both pass tests and produce robust, well-structured code. Rubric-based evaluation remains an area for further exploration: as models are deployed in production environments, qualitative assessment of code quality -- not just whether it passes tests -- becomes essential for understanding real-world capability.
\vspace{1em}

Some models write good code, scoring highly on the rubrics, but it still fails final validation, resulting in a low Pass@1. For instance, on Integration tasks, Kimi K2 Instruct achieves $75.0\%$ rubric quality on Integration tasks despite only $18.3\%$ Pass@1 -- a 56.7 percentage point gap. Its rubric breakdown shows $78.0\%$ on Functional criteria (correctly identifying service configurations) and $72.1\%$ on Robustness criteria (strong defensive coding with try-except blocks and input validation), but failures cluster around specific verification steps rather than fundamental implementation gaps. Similarly, Kimi K2.5 achieves $80.9\%$ on Functional criteria for Observability tasks but only $19.7\%$ Pass@1.
\vspace{1em}

Other models pass tests yet score poorly on the rubrics, indicating they have fragile code. Grok 4 achieves $36.3\%$ Pass@1 on Integration but only $46.4\%$ Robustness -- the lowest among models with comparable correctness. On the other hand, for Observability, six models score around 70\% on the rubrics, while Pass@1 varies from 18\% to 32\%.
\vspace{1em}

\section{Conclusion}
Performance on APEX\textendash SWE is determined by epistemic discipline, not raw coding capability. Across 200 held-out tasks spanning Integration and Observability domains, the models that succeed are those that treat incomplete information as a problem to solve rather than a gap to fill with assumptions. This distinction separates frontier performers from the rest of the field. 
\vspace{1em}

In the context of Integration tasks, epistemic discipline appears as architectural precision. The best models extract explicit specifications from task descriptions, anchor implementations on verified constraints rather than inferred ones, and perform closed-loop verification before declaring success.
\vspace{1em}

In the context of Observability tasks, epistemic discipline manifests as diagnostic agency. The best models treat log analysis as an iterative search problem, filtering noise and triangulating across multiple sources before attempting repairs. 
\vspace{1em}

Our findings suggest that progress in AI software engineering will depend less on training models to write better code, and more on teaching agents to emulate a rigorous engineering process -- gathering information systematically, verifying specifications before implementation, and refusing to declare success until empirical reality aligns with intended design.

\section{Related Work}

\paragraph{Unit-Level Code Generation} Benchmarks such as HumanEval ~\cite{chen2021humaneval} and MBPP ~\cite{austin2021mbpp} evaluate standalone function generation, but they are now largely saturated, with frontier models exceeding 90$\%$ Pass@1. Extensions include HumanEval Pro~\cite{yu2024humanevalpro} for self-invoking tasks, MultiPL-E~\cite{cassano2023multiple} for multilingual coverage, and LiveCodeBench ~\cite{jain2024livecodebench} for contamination-resistant evaluation. Still, unit-level tasks differ fundamentally from real engineering work, which requires navigating existing codebases and debugging complex system interactions.
\vspace{1em}

\paragraph{Repository-Level Code Generation} SWE-bench measures model performance on real GitHub issues that require multi-file patches. However, \citet{wang2025solved} found that 7.8$\%$ of ``passing'' patches failed developers' actual test suites, and \citet{yu2025utboost} showed that enhanced testing can significantly shift model rankings. SWE-bench Verified (OpenAI, 2025) adds human verification, while SWE-Bench Pro~\cite{deng2025swebenchproaiagents} targets similarly structured but harder tasks. Nonetheless, these benchmarks remain limited to single-repository bug fixing and exclude observability, infrastructure, and cross-service integration.
\vspace{1em}

\paragraph{Tool Orchestration and Function Calling} ComplexFuncBench~\cite{zhong2025complexfunc} evaluates multi-step function calling across booking-domain APIs. MSC-Bench~\cite{dong2025mscbench} provides a large-scale evaluation of tool orchestration within MCP ecosystems, covering 491 servers and a five-level curriculum ranging from simple retrieval to cross-server planning. BFCL v2~\cite{mao2024bfcl} and ToolHop~\cite{ye2025toolhop} offer additional perspectives. These benchmarks excel at measuring tool selection and coordination, but they stop short of evaluating infrastructure implementation. Tasks focus on calling APIs correctly rather than building production systems that persist data, deploy serverless functions, or implement full end-to-end business logic.
\vspace{1em}

\paragraph{Domain-Specific Integration} CRMArena \cite{huang2025crmarena} evaluates CRM workflows, where agents achieve below 65$\%$ success rates. ELT-Bench~\cite{jin2025eltbench} measures data pipeline construction, with the best agent correctly generating only 3.9$\%$ of data models. OSWorld~\cite{xie2024osworld} examines multimodal agents performing open-ended computer tasks across operating systems, and TheAgentCompany~\cite{xu2025theagentcompany} simulates a software company environment with about 30$\%$ autonomous task completion. These benchmarks focus on domain-specific expertise rather than the heterogeneous, cross-platform integration that characterizes real-world production engineering.
\vspace{1em}

Multi-service tasks prove substantially harder than single-service tasks. Tasks involving only one service ($n=53$) achieve $39.5\%$ mean Pass@1, while tasks requiring two or more services ($n=47$) drop to $18.6\%$ -- a 20.9 percentage point gap. The difficulty compounds: models must coordinate authentication, data transformation, and error handling across multiple APIs simultaneously.
\vspace{1em}

\section{Acknowledgments}
We thank Shubham Badgujar, Mayank Bharati, Sumit Jain, Rakshit Mandloi, Akshat Saini, and Gaurish Sood for their help on the evaluation harness and executing the evaluations.
We thank Debnil Sur, and Sarah Yun for their work on the development of the task shape.
We thank the engineers from the Mercor marketplace, as well as Pranav Aggarwal, David Bai, Marco Burstein, Priyanshu Gupta, Felix Huang, Surya Krishnapillai, Pranav Mehta, Srini Rajagopal, Prabal Sonakiya, and Gordon Su for their valuable contributions in generating the tasks. We used LLMs to assist with drafting and refinement.

\bibliography{anthology,apexcode}
\bibliographystyle{acl_natbib}

\appendix

\section{Model details}
All models are called via LiteLLM with retry logic (exponential backoff) and a maximum of three attempts per request. Model details are described in Table~\ref{tab:model_details}.

\begin{table*}[b]
\centering
\small
\caption{Information about the models tested on APEX\textendash SWE.}
\label{tab:model_details}

\begin{tabular}{l|c|ccc}
\toprule
\textbf{Model}                       & \textbf{Provider} & \textbf{Context Window (tokens)} & \textbf{Max Output} & \textbf{Thinking} \\
\midrule
Claude Opus 4.5                    & Anthropic         & $200{,}000$                     & $64{,}000$            & High              \\
Claude Opus 4.6                    & Anthropic         & $1{,}000{,}000$                 & $128{,}000$           & High              \\
Claude Sonnet 4.5                      & Anthropic         & $200{,}000$                     & $64{,}000$            & High              \\
Cognition SWE-1.6 Preview          & Cognition         & $180{,}000$                     & ---                   & High              \\
DeepSeek V3.2                        & DeepSeek AI       & $262{,}144$                         & $30{,}000$            & NA                \\
Gemini 3 Pro                 & Google            & $1{,}000{,}000$                     & $65{,}536$            & High              \\
GPT\textendash 5.1 Codex             & OpenAI            & $272{,}000$                         & $128{,}000$            & High              \\
GPT\textendash 5.2 Codex             & OpenAI            & $272{,}000$                         & $128{,}000$            & High              \\
Grok 4                               & xAI               & $256{,}000$                         & $128{,}000$            & [On by default]                \\
Kimi K2 Instruct                     & Moonshot AI       & $262{,}144$                         & $32{,}768$            & NA                \\
Kimi K2.5                            & Moonshot AI       & $262{,}144$                         & $32{,}768$            & NA                \\
\bottomrule
\end{tabular}
\end{table*}

\begin{table*}[b]
\small
\centering
\caption{Performance of models on the APEX\textendash SWE leaderboard ($n=200$) compared with the open\textendash source APEX\textendash SWE dev set ($n=50$) using combined Pass@1.}
\label{tab:apex_os_shift}
\begin{tabular}{l|cccc}
\toprule
\textbf{Model}
& \makecell{\textbf{Leaderboard} \\ \textbf{Pass@1}}
& \makecell{\textbf{Open-Source} \\ \textbf{Pass@1}}
& \makecell{\textbf{Score} \\ \textbf{Diff}}
& \makecell{\textbf{LB Rank $\rightarrow$} \\ \textbf{OS Rank}} \\
\midrule
Claude Opus 4.6 (High)              & $\mathbf{40.5\%}$ & $\mathbf{58.7\%}$ & $+18.2$ & $1 \rightarrow 1$ \\
Claude Opus 4.5 (High)              & $38.7\%$          & $57.3\%$          & $+18.6$ & $2 \rightarrow 2$ \\
Cognition SWE-1.6 Preview (High)    & $31.7\%$          & $52.7\%$          & $+21.0$ & $3 \rightarrow 3$ \\
Claude Sonnet 4.5 (High)            & $31.0\%$          & $44.0\%$          & $+13.0$ & $4 \rightarrow 5$ \\
GPT\textendash 5.2 Codex (High)     & $30.8\%$          & $40.7\%$          & $+9.9$  & $5 \rightarrow 7$ \\
Kimi K2.5                           & $30.3\%$          & $50.0\%$          & $+19.7$ & $6 \rightarrow 4$ \\
GPT\textendash 5.1 Codex (High)     & $29.0\%$          & $38.7\%$          & $+9.7$  & $7 \rightarrow 8$ \\
Gemini 3 Pro (High)                 & $26.8\%$          & $41.3\%$          & $+14.5$ & $8 \rightarrow 6$ \\
Grok 4                              & $21.0\%$          & $27.3\%$          & $+6.3$  & $9 \rightarrow 9$ \\
DeepSeek V3.2                       & $15.8\%$          & $21.3\%$          & $+5.5$  & $10 \rightarrow 10$ \\
Kimi K2 Instruct                    & $11.2\%$          & $16.0\%$          & $+4.8$  & $11 \rightarrow 11$ \\
\bottomrule
\end{tabular}
\end{table*}

\section{Comparing the APEX\textendash SWE Leaderboard with APEX\textendash SWE Open-Source}
We compare model performance on the held-out leaderboard ($n=200$) against the open-source development set ($n=50$). The leaderboard comprises 100 Integration and 100 Observability tasks; the open-source set contains 25 of each. Both sets use identical evaluation methodology, differing only in sample size. The $4\times$ larger sample size reduces variance and better reflects expected performance on novel tasks.
\vspace{1em}

Table~\ref{tab:apex_os_shift} presents Pass@1 scores -- averaging Integration and Observability performance -- for each model on both sets. Claude Opus 4.6 leads the leaderboard at $40.5\%$, followed by Claude Opus 4.5 at $38.7\%$, and both maintain their positions on the open-source set at $58.7\%$ and $57.3\%$ respectively. The bottom tier is equally stable: Grok 4, DeepSeek V3.2, and Kimi K2 Instruct hold ranks 9--11 on both sets. We see two patterns.
\vspace{1em}

First, rankings are consistent at the extremes: the top two models (Claude Opus 4.6 and 4.5) and the bottom three (Grok 4, DeepSeek V3.2, and Kimi K2 Instruct) hold identical positions on both sets. The middle tier shows more variance -- Kimi K2.5 jumps from rank 6 to rank 4 on the open-source set, while GPT\textendash 5.2 Codex drops from rank 5 to rank 7. Second, absolute scores are uniformly higher on the open-source set, averaging $40.7\%$ compared to $27.9\%$ on the leaderboard -- an inflation of 12.8 percentage points. These differences are due to the small sample size.
\vspace{1em}

\end{document}